\documentclass[final]{neus2025}

\usepackage{MnSymbol}
\usepackage{diagbox}
\usepackage{wrapfig}
\usepackage{makecell}
\usepackage[nooneline]{caption} 
\usepackage{adjustbox}

\title[]{State-Dependent Conformal Perception Bounds for\\ Neuro-Symbolic Verification of Autonomous Systems}
\usepackage{times}

\coltauthor{%
\Name{Thomas Waite} \Email{waitet@rpi.edu} 
\\ 
 \addr Rensselaer Polytechnic Institute, Troy, New York \\
 \Name{Yuang Geng} \Email{yuang.geng@ufl.edu}\\
 \Name{Trevor Turnquist} \Email{trevor.turnquist@ufl.edu}\\
 \Name{Ivan Ruchkin}\thanks{Co-last authors: equal contribution in supervision.} 
 \Email{iruchkin@ece.ufl.edu}\\
 \addr University of Florida, Gainesville, Florida \\
\Name{Radoslav Ivanov}\footnotemark[1] \Email{ivanor@rpi.edu}\\
\addr Rensselaer Polytechnic Institute, Troy, New York
}

\newtheorem{problem}{Problem}
\newtheorem*{remarkstar}{Remark}

\begin{document}

\maketitle

\begin{abstract}
It remains a challenge to provide safety guarantees for autonomous systems with neural perception and control. A typical approach obtains symbolic bounds on perception error (e.g., using conformal prediction) and performs verification under these bounds. However, these bounds can lead to drastic conservatism in the resulting end-to-end safety guarantee. This paper proposes an approach to synthesize symbolic perception error bounds that serve as an optimal interface between perception performance and control verification. The key idea is to consider our error bounds to be heteroskedastic with respect to the system's state --- not time like in previous approaches. These bounds can be obtained with two gradient-free optimization algorithms. We demonstrate that our bounds lead to tighter safety guarantees than the state-of-the-art in a case study on a mountain car. 
\end{abstract}

\begin{keywords}
Neural network verification, conformal prediction, gradient-free optimization
\end{keywords}

\section{Introduction}

\looseness=-1
Modern autonomous systems such as \cite{waymo}'s self-driving cars and \cite{volocity}'s air taxis show impressive capabilities of neural perception and control, but providing safety guarantees on such systems remains difficult. The primary obstacle in verifying safety is that the agents operate in complex, stochastic environments perceived through high-dimensional measurements. 
Purely formal (symbolic) verification techniques require realistic environmental models to make the verification result meaningful in practice. Environmental models can be constructed from first principles (e.g. via a pinhole camera model or by tracing LiDAR rays), but such models cannot easily account for unexpected stochastic complexities of real systems such as LiDAR reflections (\cite{ivanov20a}). Alternatively, environmental models could be learned via a generative network (\cite{katz22}), but verification tools are difficult to scale to the individual pixel level, particularly for closed-loop systems (\cite{everett2021neural}). Besides, the real-world validity of such neural models remains in question. At the other extreme, purely statistical verification approaches are appealing because they capture statistical uncertainty --- but cannot exploit the knowledge of the underlying system dynamics. 

\looseness=-1
Many popular methods combine statistical and symbolic safety verification of systems with neural perception. The general approach follows two steps: first, obtain high-confidence bounds on uncertain quantities (e.g., neural perception error) using a statistical tool such as conformal prediction (CP); second, compute high-confidence reachable sets using a symbolic description of the dynamics  (\cite{lin2024verification}, \cite{muthali2023multi}, \cite{jafarpour2024probabilistic}, \cite{geng2024bridging}). A persistent challenge for these approaches, however, is \emph{overly conservative reachable sets}, particularly over long time horizons. 
In response, many methods aim to reduce conservatism in CP bounds for a variety of settings (\cite{romano2019conformalized}, \cite{sharma2024pac}, \cite{kiyani2024length}, \cite{tumu2024multi}). One particular approach by \cite{cleaveland2024conformal} reformulates the weighted conformity scores to optimize for tighter bounds than the point-wise ones by \cite{lindemann2023safe}. In effect, it exploits the fact that the conformal errors are \emph{heteroskedastic} over time. We observe that in practice, however, error is often highly correlated with state (e.g., motion blur is increased at higher speeds). Our key insight is that neural perception errors are \emph{heteroskedastic with respect to state}, and this heteroskedasticity can be utilized to reduce conservatism in symbolic reachability analysis.

In this work, we introduce \textbf{state-dependent conformal bounds} for neural perception error as an ``interface'' between neural perception and symbolic verification. To do this, we propose two methods to 
partition the state space into regions via gradient-free optimization methods. The regions are optimized such that the regional perception errors contribute minimally to over-approximation error in the symbolic reachability calculation. Our approach balances the number of regions (which can decay our guarantee due to the union bound) and the size of the error in each region. 

The proposed neuro-symbolic verification method opens the path to a new level of assurance for autonomous systems. The high-level approach is to use symbolic techniques for well-understood parts of the system (e.g.,~dynamics) and data-driven methods for high-dimensional and hard-to-model aspects (e.g.,~perception). Specifically, we abstract the perception model and obtain high-confidence data-driven bounds on the abstracted system, which are then used to construct tight, high-confidence reachable sets using a symbolic verification tool by leveraging the system dynamics. The ultimate output of our approach is a safety guarantee that provably holds with a user-specified probability. Our contributions are as follows:
\begin{itemize}
    \item A framework for providing statistical safety guarantees on neural perception and control systems that exploits heteroskedasticity in perception error over the state space. 
    \item A method for finding state-dependent neural perception error bounds via conformal prediction. The bounds and regions are selected with gradient-free optimization methods to reduce overapproximation error in symbolic high-confidence reachability computations. 
    \item A case study on mountain car demonstrating our conformal bounds lead to significantly smaller reachable sets than the state-of-the-art time-based conformal prediction methods.
\end{itemize}

\paragraph{Related Work}
Conformal prediction, originally introduced by 
\cite{vovk2005algorithmic} and \cite{shafer2008tutorial}, is an increasingly popular method for obtaining data-driven uncertainty bounds. As conformal prediction has expanded to a variety of applications, safety of autonomous systems has gained particular attention with recent examples including safe motion planning (\cite{lindemann2023safe}), safe controller design (\cite{yang2023safe}), online safety monitoring (\cite{zhang2024bayesian}, \cite{zhao2024robust}), and integration into safety decision-making frameworks (\cite{lekeufack2024conformal}). We refer the reader to \cite{lindemann2024formal} and \cite{angelopoulos2023conformal} for detailed tutorials and a broader overview of the conformal prediction field.

Deterministic reachability for neural control systems is a well-developed area. Open-loop methods focus on verifying input-output properties of networks (\cite{wang2021beta, dutta18, katz17, tran20}), while closed-loop methods interleave neural networks with symbolic dynamics to calculate reachable sets (\cite{ivanov19, dutta2019reachability, huang2019reachnn,fan2020reachnn, wang2023polar}).Recently, \cite{chakraborty2023discovering} used reachability analysis on image-controlled systems to discover unsafe initial sets, but it requires exhaustive querying of a simulator's perception map for all states, preventing analysis of real systems. 

\looseness=-1
In this work, we combine conformal prediction and neural network reachability to consider high-probability reachable sets for neural perception and control systems with known dynamics. While methods exist for reachability of stochastic systems with known or unknown dynamics (\cite{abate2007computational, abate2008probabilistic, lin2023generating, alanwar2023data, bortolussi2014statistical}), they do not consider neural components for perception or control. One notable approach from \cite{hashemi2024statistical} combines neural network reachability and conformal predictions but focuses on high-confidence reachability for systems with \emph{unknown dynamics}.

\section{Problem Formulation}
\label{sec:problem}
We consider dynamical systems with perception of the form
\begin{align}
\label{eq:system_model}
\begin{split}
    x_{k+1} = f(x_k, u_k); \quad z_k = p(x_k); \quad y_k = nn(z_k) := g(x_k) + v_k; \quad u_k = h(y_k),
\end{split}
\end{align}
where $x_k \in \mathcal{X} \subset \mathbb{R}^n$ are the system states (e.g.,~position, velocity); $z_k \in \mathbb{R}^{m_z}$ are the measurements (e.g.,~camera images) generated from an unknown perception map $p$; $y_k \in \mathbb{R}^{m_y}$ are the outputs of a neural component $nn$ trained to extract a desired function $g$ of the states from images (e.g.,~state estimates); $v_k$ is the unknown random noise introduced by the neural component $nn$; $f$ is the known plant dynamics model, and $h$ is a known controller.\footnote{While the proposed framework can handle the more general setting with the dynamics noise, for simplicity we assume no dynamics noise in the problem statement.} 

Our reasoning for this model choice is as follows. First, note that~\eqref{eq:system_model} models a standard system with neural perception where the neural component is trained to extract a low-dimensional symbolic representation of the measurements (e.g.,~car location within the lane). As discussed in prior work by~\cite{dean20}, this formulation enables a separation-principle-like control design where the controller can be developed specifically for $g$ (e.g., a linear measurement), while being robust to high-probability bounds on $v_k$ (and thus abstracting away the unknown and complex map $p$). Similarly, this formulation enables a high-confidence verification approach that abstracts away $p$ and verifies safety for the entire system, as long as high-confidence bounds on $v_k$ are known. Thus, in the remainder of the paper, we will only focus on the following \emph{abstracted system}: 
\begin{align}
\label{eq:abst_system_model}
\begin{split}
    x_{k+1} = f(x_k, u_k); \quad y_k = g(x_k) + v_k; \quad  u_k = h(y_k).
\end{split}
\end{align}
%
%
%

\paragraph{Background: reachability analysis.} Consider a system such as the one in~\eqref{eq:abst_system_model}, where we are given a known initial set $\mathcal{X}_0$ and known bounds on the noise $\|v_k\| \le b$. Reachability analysis aims to calculate reachable sets $\mathcal{X}_1, \dots, \mathcal{X}_T$ that are guaranteed to contain the state $x_k$ at each time $k$ (e.g.,~so as to verify that no unsafe states are reached). The reachable sets are typically conservatively approximated using computationally convenient shapes such as ellipsoids (\cite{althoff15}) or Taylor models (\cite{chen12}). Unfortunately, worst-case bounds on $v_k$ in~\eqref{eq:abst_system_model} may be impossible to obtain without strong and often unrealistic assumptions. 
%
Thus, the problem considered in this paper is to compute reachable sets that hold \textit{with high probability} over random initial conditions and noise trajectories. These reachable sets are useful in a number of ways, e.g., for high-confidence pre-deployment guarantees, online monitoring, or planning around other agents. 

To obtain high-confidence reachable sets, we assume we are given a dataset of $N$ trajectories ${D = \{(x_{1,0:T}, y_{1,0:T}) \dots, (x_{N,0:T}, y_{N,0:T})\}}$, where ${x_{i,0:T} = (x_{i,0}, \dots, x_{i,T})}$ is the full trajectory $i$ of states (same for measurements). To keep our notation simple, we assume all trajectories have the same length. We also assume all trajectories are generated using controller $h$ to ensure that they are on-policy and independently identically distributed (IID), such that each $x_0 \sim \mathcal{D}_0$ is sampled from an unknown distribution $\mathcal{D}_0$ over a known set $\mathcal{X}_0$. Finally, we also assume $v_k \sim \mathcal{V}_{k|k-1}$, i.e.,~the noise at each step is sampled from an unknown conditional distribution, $\mathcal{V}_{k|k-1}$, given the previous noise values and the initial state. As part of future work discussed in Section~\ref{sec:future}, we will investigate the off-policy problem where trajectories are generated using an exploration controller.

\begin{problem}[High-Confidence Reachability]
\label{prob:reachability}
Given the system in~\eqref{eq:abst_system_model}, 
a confidence level $\alpha$, and a calibration dataset of trajectories $D$, the goal is to construct a \emph{sequence of reachable sets} $\mathcal{X}_1, \dots, \mathcal{X}_T$ such that ${\mathbb{P}_{x_0 \sim \mathcal{D}_0, v_k \sim \mathcal{V}_{k|k-1}}[\forall k = 0..T: x_k \in \mathcal{X}_k] \ge 1 - \alpha}$.
\end{problem}

While Problem~\ref{prob:reachability} can be solved using existing (purely statistical) time-series conformal prediction, e.g., by~\cite{lindemann2023safe}, the resulting sets would inevitably be conservative without any knowledge of system dynamics.
The main benefit of knowing the dynamics model $f$ is that it allows us to use \textit{reachability analysis} to solve Problem~\ref{prob:reachability}, e.g.,~the authors' tool Verisig (\cite{ivanov19}) --- as long as high-confidence bounds on perception noise $v_k$ are available. 


\looseness=-1
\paragraph{Background: scalar conformal prediction.}
In the scalar CP setting by~\cite{angelopoulos2023conformal}, we are given a calibration dataset $D = \{z_1, \dots, z_N\}$, where the $z_i$ are realizations of exchangeable random variables $Z_1, \dots, Z_N$, i.e.,~${\mathbb{P}[Z_{q(1)} \le \dots \le Z_{q(N)}] = \mathbb{P}[Z_{r(1)} \le \dots \le Z_{r(N)}]}$ for any two re-ordering functions $q$ and $r$. Consider a new random variable $Z_{test}$ that is also exchangeable with the $Z_i$. Assuming the $z_i$ are sorted in increasing order, one can show that ${\mathbb{P}[Z_{test} \le z_{\lceil (N+1)(1-\alpha)\rceil}] \ge 1 - \alpha}$. In other words, the (normalized) $1-\alpha$ quantile, denoted by $\text{Quantile}(D, 1-\alpha)$, is a high-confidence upper bound on a new exchangeable sample. 

\paragraph{Background: conformal prediction for time-series data.} The time-series CP setting, e.g.,~as considered by~\cite{cleaveland2024conformal}, is more challenging. Here, $D = \{z_{1,0:T}, \dots, z_{N,0:T}\}$, and the problem is to design a bound function $\eta$ such that ${\mathbb{P}[\forall k = 0..T: Z_{test,k} \le \eta(k)]} \ge 1-\alpha$, i.e.,~the probability that the entire trajectory $Z_{test}$ is within the $\eta$ bounds is at least $1-\alpha$. A straightforward solution is to apply the scalar bounds for each time step and then obtain trajectory-wide guarantees using the probability union bound; however, this approach results in overly conservative confidence. To overcome this challenge, researchers, e.g.,~\cite{cleaveland2024conformal} and~\cite{angelopoulos2023conformal_pid}, have proposed to re-weigh the bounds $\eta(k)$ at each time step $k$ to tighten up the bounds. 

\paragraph{Novel setting: state-dependent conformal prediction.} As noted above, we aim to obtain high-confidence bounds on perception noise $v_k$ that would enable reachability analysis as a solution to Problem~\ref{prob:reachability}. Although such bounds on $v_k$ can be directly obtained using time-series conformal prediction, they tend to be conservative: existing works reduce conservativeness by exploiting heteroskedasticity (i.e., varied uncertainty in $v_k$) \textit{over time}. In contrast, we put forward a more effective approach to exploit heteroskedasticity \textit{over the state space}. Since perception error likely varies drastically within the state space, we expect state-dependent bounds to separate high-noise from low-noise regions and result in much tighter reachable sets.

To be precise, we propose to partition the state space into $M$ disjoint regions: ${\mathcal{X} = \mathcal{S}_1 \cupdot  \cdots \cupdot  \mathcal{S}_M}$. Each state $x$ will correspond to a piecewise-constant perception error bound $\eta(x)$ 
determined by the region $\mathcal{S}_i \ni x$, where $\eta$ will be chosen to satisfy a high-confidence guarantee on the noise within each region: ${\mathbb{P}[\forall k = 0..T: (x_k \in \mathcal{S}_i \Rightarrow \|v_k\| \le \eta(x_k))] \ge 1 - (\alpha/M)}$. The per-region guarantees would allow us to maintain the overall high-confidence guarantee
from Problem~\ref{prob:reachability} (proved in Section~\ref{sec:approach}), with the added benefit that the final reachable sets, $\mathcal{X}_i$, may be much tighter than those obtained through time-series CP. Of course, this approach requires a suitable partitioning of the state space, which is the main problem of this paper.

\begin{problem}[State-Dependent Conformal Prediction]
\label{prob:state_dependent_cp}
Given the system in~\eqref{eq:abst_system_model}, 
a confidence level $\alpha$, and a calibration dataset $D$, the goal is to \emph{partition the state space} into $M$ disjoint regions, i.e., $\mathcal{X} = \mathcal{S}_1 \cupdot \cdots \cupdot \mathcal{S}_M$ and compute a corresponding \emph{noise bound function} $\eta(x)$, so as to minimize a loss function $\mathcal{L}(D, \mathcal{S}_1, \dots, \mathcal{S}_M)$ correlated with tighter reachable sets $\mathcal{X}_i$, as defined in Problem~\ref{prob:reachability}.
\end{problem}

\begin{remarkstar}
Problem~\ref{prob:state_dependent_cp} has two parts: 1) identifying a suitable loss function $\mathcal{L}$ and 2)~solving the resulting optimization problem. Both of these parts are the main contributions of this paper.
\end{remarkstar}
\section{Approach}
\label{sec:approach}
This section presents the proposed approach, starting with the theoretical results that demonstrate its soundness and followed by the algorithms to partition the state space and compute reachable tubes. 

\begin{figure}[t]
\captionsetup{justification=centering}
    \centering
    \includegraphics[trim={9mm 1mm 10mm 1mm},clip,width=\textwidth]{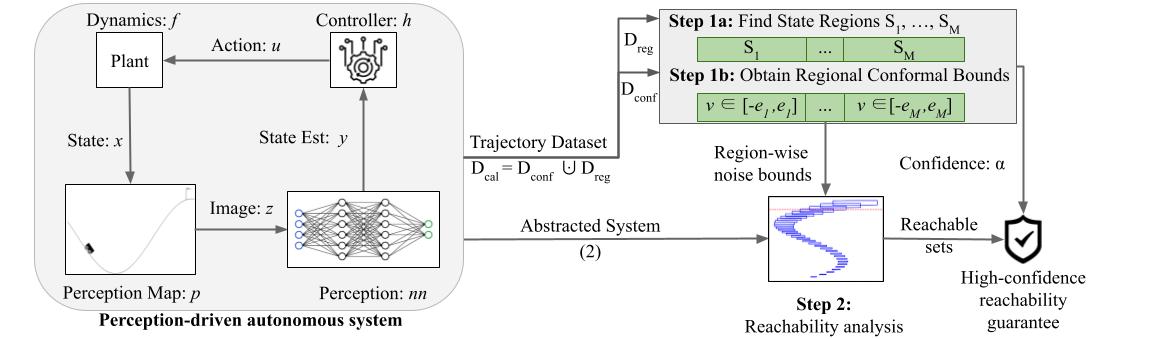}
    \vspace{-7mm}
    \caption{An overview of our approach to high-confidence reachability. }
    \label{fig:approach-overview}
    \vspace{-5mm}
\end{figure}

\subsection{Approach Overview}

\looseness=-1
At a high level, the proposed approach consists of two steps (corresponding to Problems~\ref{prob:state_dependent_cp} and~\ref{prob:reachability}, respectively), illustrated in Figure~\ref{fig:approach-overview}. Step 1 is to partition the state space $\mathcal{X} = \mathcal{S}_1 \cupdot \dots \cupdot \mathcal{S}_M$ and design a region-based function $\eta(x)$ such that ${\mathbb{P}_{x_0\sim \mathcal{D}_0, v_k \sim \mathcal{V}_{k|k-1}}[\forall k = 0..T: \|y_k - g(x_k)\| \le \eta(x_k)]} \ge 1-\alpha$, where $\mathcal{D}_0$ and $\mathcal{V}_{k|k-1}$ are unknown, but $\mathcal{D}_0$ is assumed to have support over a known set $\mathcal{X}_0$. Step 2 is to perform worst-case (deterministic) reachability analysis of the abstracted system in~\eqref{eq:abst_system_model} 
from initial set $\mathcal{X}_0$. Here, $v_k$ is treated as bounded noise with state-dependent bounds $\eta(x)$. The rest of this subsection shows that the condition in Step~1 ensures the reachable sets in Step~2 solve Problem~\ref{prob:reachability}. The following subsections provide a specific approach for each step, leading to tight reachable sets.

\paragraph{Step 1: region-based perception noise bounds.} Before we discuss how to partition the state space (in Section~\ref{sec:state_regions}), we first outline the requirements for this partition. In particular, we aim to apply a union bound over all regions, so each region must satisfy the following upper bound on perception error: ${\mathbb{P}_{x_0\sim \mathcal{D}_0, v_k \sim \mathcal{V}_{k|k-1}}[\exists k = 0..T: (x_k \in \mathcal{S}_i \wedge \|y_k - g(x_k)\| > \eta(x_k))] \le (\alpha/M)}$.\footnote{It is possible to require that different regions have different confidence bounds, as long as the overall guarantee of $1-\alpha$ is reached. For simplicity, however, we require all regions to have the same $1 - (\alpha/M)$ guarantee.} The next proposition shows how this guarantee leads to the overall guarantee over the entire state space. All proofs are provided in the appendix.

\begin{proposition}
\label{prop:sd_conformal_prediction}
Consider the abstracted system in~\eqref{eq:abst_system_model} with the state space partitioned into $M$ regions, $\mathcal{X} = \mathcal{S}_1 \cupdot \dots \cupdot \mathcal{S}_M$, and assume a high-confidence bound within each region: 
\vspace{-5px}
\begin{align*}
\mathbb{P}_{x_0\sim \mathcal{D}_0, v_k \sim \mathcal{V}_{k|k-1}}[\exists k = 0..T: (\|y_k - g(x_k)\| > \eta(x_k) \wedge x_k \in \mathcal{S}_i)] \le \frac{\alpha}{M}.    
\end{align*}
\vspace{-10px}
$\\$
Then the trajectory-wide bound holds:  ${\mathbb{P}_{x_0\sim \mathcal{D}_0, v_k \sim \mathcal{V}_{k|k-1}}[\exists k = 0..T: \|y_k - g(x_k)\| > \eta(x_k)]} \le \alpha$.
\end{proposition}
%

\paragraph{Step 2: High-confidence reachability analysis.} Given the trajectory-wide guarantee on $v_k$, the following theorem shows that performing worst-case reachability analysis using $\eta$ bounds on $v_k$ will produce reachable sets that satisfy the condition in Problem~\ref{prob:reachability}.

\begin{theorem}
\label{thm:reachability}
Consider the abstracted system in~\eqref{eq:abst_system_model},
with $x_0$ sampled from an unknown distribution $\mathcal{D}_0$ with support over a known set $\mathcal{X}_0$. Suppose we are given a bound function $\eta$ such that $\mathbb{P}_{x_0\sim \mathcal{D}_0, v_k \sim \mathcal{V}_{k|k-1}}[\exists k = 0..T: \|y_k - g(x_k)\| > \eta(x_k)] \le \alpha$. Suppose worst-case reachable sets $\mathcal{X}_1, \dots, \mathcal{X}_T$ are computed for~\eqref{eq:abst_system_model}, with initial set $\mathcal{X}_0$ and noise bounds ${\|v_k\| \le \eta(x_k)}$. Then:
\vspace{-5px}
\begin{align*}
    \mathbb{P}_{x_0\sim \mathcal{D}_0, v_k \sim \mathcal{V}_{k|k-1}}[\forall k = 0..T: x_k \in \mathcal{X}_k] \ge 1 - \alpha.
\end{align*}
\end{theorem}

\subsection{State-Dependent Conformal Prediction}
\label{sec:state_regions} 

This subsection presents an optimization-based approach for partitioning the state space into regions $\mathcal{X} = \mathcal{S}_1 \cupdot \dots \cupdot \mathcal{S}_M$ that 1)~satisfy the condition in Proposition~\ref{prop:sd_conformal_prediction} and 2)~reduce the approximation error incurred by the subsequent reachability task. We first describe how to calculate trajectory-wide guarantees per region for \textit{any} partition. 
We define \emph{region-specific trajectory subsets}: 
$$D_{\mathcal{S}_i} = \{(x_{1,m_1:n_1}, y_{1,m_1:n_1}), \dots, (x_{N,m_N:n_N}, y_{N,m_N:n_N}) \mid x_{i,j} \in \mathcal{S}_i\},$$
where we consider the sub-trajectory of each $x_{i,0:T}$ which is contained in $\mathcal{S}_i$; note that the time steps in each sub-trajectory are the same as in the full one. Given $D_{\mathcal{S}_i}$, the \emph{sub-trajectory non-conformity scores} are defined as the maximum sub-trajectory-wide perception error within each region,
  \begin{align}
     \delta^j_{\mathcal{S}_i} = \max_{t=m_j..n_j}\|x_{j, t} - y_{j, t}\| \text{ for } j = 1..N;    \text{ and } \delta^{N+1}_{\mathcal{S}_i} = \infty 
 \end{align}

Next, we apply scalar conformal prediction to these non-conformity scores to obtain the corresponding perception error confident bound for each region for Proposition~\ref{prop:sd_conformal_prediction}.

\begin{proposition}[High-confidence region-based perception error bound]
\label{prop:eta}
\looseness=-1
Given a confidence level $\alpha$ and sub-trajectory error dataset $\Delta_i = \{\delta^1_{\mathcal{S}_i}, \dots, \delta^{N+1}_{\mathcal{S}_i}\}$ for each region $\mathcal{\mathcal{S}_i}_i$, the following \emph{perception error bound} $\eta(x)$ satisfies the conditions in Proposition~\ref{prop:sd_conformal_prediction}:
%
\begin{align*}
\eta(x) = \text{\emph{Quantile}}\left(\Delta_i, 1-\frac{\alpha}{M}\right) \text{ if } x \in \mathcal{S}_i.
\end{align*}
\end{proposition}


For conformal perception bounds in each region, two loss functions are applied to optimize the partitioning: \emph{Experience Loss (EL)} and \emph{Experience Time-Decay Loss (ETDL)}. EL is designed to prioritize frequently visited regions by assigning higher weights. This strategy tightens conformal error bounds in these areas, under the assumption that they are of greater significance for the reachability analysis since we need to inflate reachable sets more frequently. Less frequently visited regions receive lower weights to balance the optimization process. EL is defined as:
\vspace{-5px}
\begin{equation}
    \mathcal{L}_{EL} = \sum_{i=1}^M\sum_{x_{j,t} \in D_{\mathcal{S}_i}} w_i \eta(x_{j,t}),
\text{ where the weights are } w_i = \frac{|D_{\mathcal{S}_i}|}{\sum_{j=1}^{M} |D_{\mathcal{S}_j}|}.
\end{equation}
\vspace{-10px}
$\\$
ETDL extends EL by incorporating a time-decay weighting strategy. Since over-approximation errors tend to accumulate over time, ETDL assigns larger weights to states earlier in the trajectory. This helps avoid accumulating error early during verification. ETDL is defined below:
\vspace{-5px}
\begin{equation}
    \mathcal{L}_{ETDL} = \sum_{i=1}^{M} \sum_{x_{j,t} \in D_{\mathcal{S}_i}} w_i  \lambda_t  \eta(x_{j,t}),
\end{equation}
\vspace{-10px}
$\\$
where $\lambda_t$ is a decreasing, time-dependent weight function; we use exponential decay in our experiments: $\lambda_t = 0.9^t$. 
%
We are now ready to state the region-based optimization problem considered in this paper.
\vspace{-5px}
\begin{definition}[Reachability-Informed Region Optimization]
Given a calibration dataset of trajectories $D$ and a confidence bound $\alpha$, \emph{the reachability-informed region optimization problem} is to select $M$ regions that: 
\end{definition}
\vspace{-6mm}
\begin{align}
    \label{eq:optimziation_cp_region}
    \begin{split}
    \min_{\mathcal{S}_1, \dots, \mathcal{S}_{M}}&\ \mathcal{L}, ~ \text{where} ~ \mathcal{L} =   
    \begin{cases}
        \mathcal{L}_{EL}, & \text{if EL is chosen} \\
        \mathcal{L}_{ETDL}, & \text{if ETDL is chosen}
    \end{cases}\\
    \text{s.t.} &\ \mathcal{X} = \mathcal{S}_1 \cupdot \dots \cupdot \mathcal{S}_M, \\
    &\ \Delta_i = \{\delta^1_{\mathcal{S}_i}, \dots, \delta^{N+1}_{\mathcal{S}_i}\}, i = 1, \dots, M, \text{ and }\\
    &\ \ \eta(x) = \text{Quantile}\left(\Delta_i, 1-\frac{\alpha}{M}\right) \text{ if } x \in \mathcal{S}_i, i = 1, \dots, M.\\
    \end{split}
\end{align}

\paragraph{Solving the optimization problem.} The problem in~\eqref{eq:optimziation_cp_region} is ill-defined for arbitrary shapes for $\mathcal{S}_i$. As a first step, we consider boxes for the regions and two \textit{gradient-free global search algorithms}: Genetic Algorithm (GA) (\cite{mirjalili2019}) and Simulated Annealing (SA) (\cite{vanlaarhoven1987}). Both methods are well-suited for this problem because they do not require gradient information and
can effectively search for globally optimal partitions of the state space. GA explores the solution
through evolutionary computation: selection, crossover, and mutation to refine candidate region solutions iteratively. In contrast, SA operates through stochastic perturbations of regions, using a probabilistic criterion to escape local minima while gradually converging on an optimal solution.

\subsection{Reachable Tube Computation}
\label{subsec:reachability}
Given the state region partitions $\mathcal{S}_1, \dots, \mathcal{S}_M$ and perception bound function $\eta$, we have now obtained high-confidence bounds for the unknown random noise $v_k$ in the abstracted system from~\eqref{eq:abst_system_model}. For each region, we define the \textit{conformal error bound} explicitly: $e_i = \eta(x_k)$ if $x_k\in\mathcal{S}_i$ for $i=1,\dots,M$. From Theorem~\ref{thm:reachability}, in the case of $L_\infty$ norm bounds on $v_k$, we have that $v_k \in [-e_i, e_i]$. Thus, the system for which we need to compute reachable sets becomes: 
\begin{align}
\label{eq:abst_verification_model}
\begin{split}
    x_{k+1} = f(x_k, u_k); \quad 
    y_k = g(x_k) + [-e_i,e_i] ~~ \text{ if } x_k \in S_i; \quad
    u_k = h(y_k).
\end{split}
\end{align}
%
To compute high-confidence reachable set analysis for our abstracted dynamical system as defined in Problem \ref{prob:reachability}, we can use any reachability tool for hybrid systems. One such tool is the authors' tool Verisig (\cite{ivanov2021b}). To encode the regional perception bounds in a hybrid system, we add transitions between the plant $f$ and controller $h$ with  guards determined by the regions ($x_k\in \mathcal{S}_i$) and resets that inflate the measurement model $g(x_k)$ with the corresponding error interval $[-e_i, e_i]$.

Note that introducing additional transitions to a system can make scalability challenging: reachable sets that intersect with multiple regions must be considered separately. This leads to longer verification time as each ``branch" must be verified separately. In highly-branching verification tasks, individual branches are often strict subsets of other branches with larger reachable sets, leading to redundant verification. Next, we describe our method to remove this redundancy in Verisig. 

In Verisig, reachable sets are represented with \emph{Taylor Models} (TMs), introduced by~\cite{makino03}. Informally, a TM encloses a function $f$ over a specified domain. Formally, a TM for a function $f$ is an over-approximation for $f$ containing a polynomial $p_f$ and worst-case error bound $I_f$ for a given domain $D$, such that $f(x) \in \{p_f(x) + e \mid e \in I_f\}~\forall x\in D$. To remove redundant branches, we aim to identify when one TM ``parent" branch encloses another. In general, this is not trivial because TM ranges are evaluated via interval arithmetic and produce boxes from their symbolic and error components. Thus, a conservative method to check for inclusion is to transform the ``parent'' into a box and check whether a conservative approximation of the ``child'' is fully within this box. While this method may introduce additional error due to transforming ``parent'' TMs, Verisig already implements \emph{shrink-wrapping} of TMs to reduce long-term over-approximation whenever remainders grow large (\cite{ivanov2021b}). Shrink-wrapping resets TMs to be fully-symbolic and contain their original range with no remainder. Thus, we opportunistically check for redundant subset branches whenever a branch is shrink-wrapped. Any subset branches of a newly shrink-wrapped branch are removed from the verification, thus enhancing its scalability.


\section{Case Study: Mountain Car}
\label{case_study}
\looseness=-1
We evaluate the proposed neuro-symbolic verification method on Mountain Car (MC), a popular reinforcement learning benchmark from OpenAI's \cite{mc}. Consistently throughout, we measure our results against the time-series approach from \cite{cleaveland2024conformal} and refer to this approach as the ``baseline". As the baseline requires a fixed time horizon, we use $T=90$. In all methods, we set $\alpha = 0.05$ so that computed reachable sets contain the real trajectory with 95\% confidence.
\begin{figure}[h]
  \begin{center}
  \subfigure[Perception Errors for all 4,000 \newline trajectories in $D$ showing  hetero- \newline skedasticity over time and state.][b]{
  \label{fig:perception_errors}
          \includegraphics[trim={1mm 0mm 5mm 7mm},clip,width=0.45\textwidth]{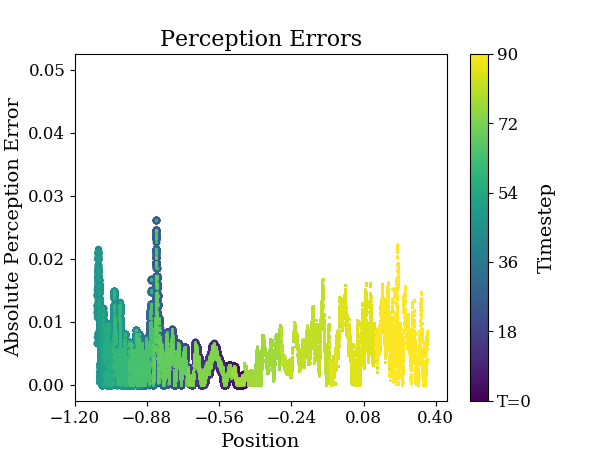}
          }
    \subfigure[State-based regional conformal bounds for $M=3$ regions vs. the baseline bounds.
    ][b]{
        \label{fig:conformal_bound_comp}
          \includegraphics[trim={1mm 0mm 5mm 7mm},clip,width=0.45\textwidth]{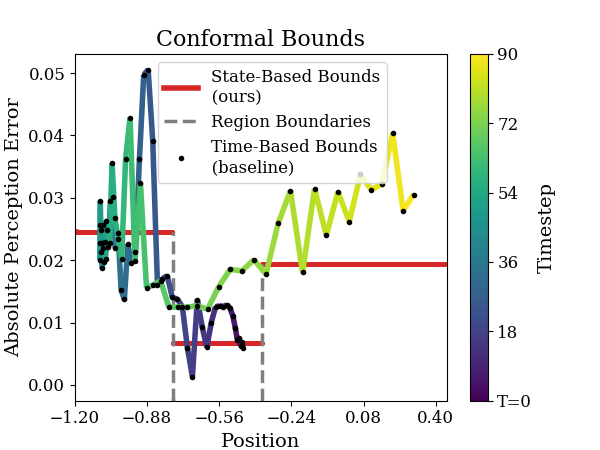}.}
  \vspace{-2mm}
  \caption{Perception errors and their respective bounds under our method and time-based baseline.} 
  \label{fig:state_time_errs}
  \end{center}
  \vspace{-5mm}
  \vspace{-4mm}
\end{figure}

\begin{wrapfigure}[18]{H}{0.33\textwidth}
\vspace{-22mm}
  \centering
    \includegraphics[trim={1mm 1mm 5mm 7mm},clip,width=0.33\textwidth]{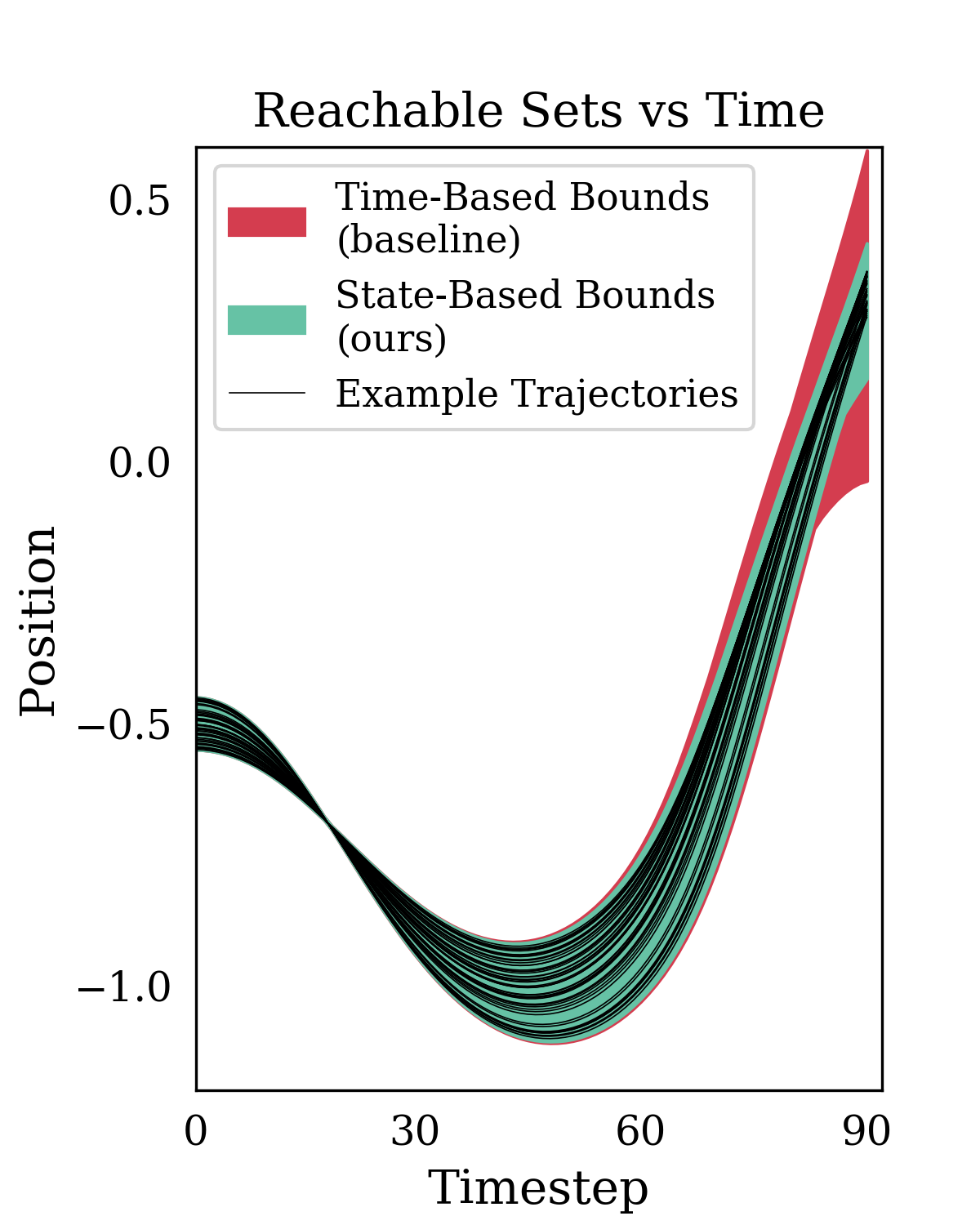}
  \caption{ Reachable Sets for our state-based conformal perception bounds (GA+ETDL, $M=7$) vs. the time-based baseline bounds. 
  Our reachable sets are tighter than the time-based method \emph{at all timesteps}.}
  \label{fig:reachtubes}
  \end{wrapfigure}

\looseness=-1
\vspace{-2.5pt}
\paragraph{Control and Perception Models.}
Per Section 
\ref{sec:problem}, we consider a modular control pipeline with a low dimensional (i.e. state-based) neural controller and a perception model that extracts low-dimensional representations from high-dimensional observations. In particular, we use a controller $h$ adopted from \cite{ivanov2020verifying} that was pre-trained and pre-verified to be safe (i.e., reaching the top of the hill with a reward of at least 90) when observing the ground-truth position and velocity starting from the initial set $p_0\in[-0.55, -0.45]$. For the perception model, we use a state estimator $nn$ that predicts the position $x$ of the car using a gray-scaled image from the MC simulator. Since we cannot produce velocity estimates from single images, we provide the controller with ground truth velocity and leave multi-image perception for future work. See Appendix \ref{sec:MC_appendix} for details. 
\vspace{-2.5pt}
\paragraph{Data Collection.}
We generated a dataset $D$ of 4,000 trajectories by simulating MC with the perception model $nn$ and controller $h$ described above. The initial states $\mathcal{X}_0$ are sampled uniformly from $[-0.55, -0.45]$, and trajectories are terminated after reaching the goal state $x=0.45$. We emphasize that while the perception model was trained on pre-deployment data with contrast noise, the dataset $D$ was generated by adding blur noise to image observations so as to demonstrate that the proposed method can handle out-of-distribution deployment noise on the perception model (details in Appendix \ref{sec:MC_appendix}). The perception errors for this dataset are shown in Figure \ref{fig:perception_errors} -- note the drastic heteroskedasticity over the state space exposed by the added blur noise. We split $D$ evenly into 2,000-trajectory calibration and test sets, $D_{cal}$ and $D_{test}$, respectively. $D_{cal}$ is further split into two disjoint sets: $D_{reg}$ for determining region edges via \eqref{eq:optimziation_cp_region} and $D_{conf}$ for finding the regional conformal bounds. 
$D_{test}$ is reserved for testing the conservativeness of the probabilistic guarantees. 
\looseness=-1
\vspace{-2.5pt}
\paragraph{Conformal Bound Computations.} We compute regions and regional conformal bounds in three ways. First, we use our state-based method with all combinations of optimization algorithms \{SA, GA\}, loss functions \{EL, ETDL\}, and regions $M \in \{2,\dots,7\}$. To solve for the regions via \eqref{eq:optimziation_cp_region}, we randomly select 500 trajectories from $D_{cal}$ and use the remaining 1,500 trajectories to find conformal bounds. Second, as an ablation, we compute conformal bounds based on partitioning the state space \emph{uniformly} into $M = \{1,\dots,7\}$ equally sized regions, using all 2,000 trajectories in $D_{cal}$ for conformal bounds (as we 
do not need to synthesize regions). 
Third, we compute time-based conformal bounds for the baseline comparison. Using the algorithm described by \cite{cleaveland2024conformal}, we randomly select 100 trajectories from $D_{cal}$ to set the $\alpha$ values and use the remaining 1,900 to set the conformal bounds. Figure~\ref{fig:conformal_bound_comp} illustrates regions and conformal bounds for GA+ETDL ($M=3$) and the time-based conformal bounds for an example trajectory. 
\vspace{-2.5pt}
\paragraph{Reachable Set Size Evaluation.} Table \ref{tab:verify_results} summarizes the average verification time and maximal reachable set sizes under each experiment.
To compute reachable sets for our state-based methods, we follow the approach described in Section \ref{subsec:reachability} and encode the regional perception errors in Verisig with $M$ discrete jumps between the dynamics and controller, corresponding to each region. For the time-based method, a different perception error bound is used at each time step, as per the bounds shown in Figure~\ref{fig:conformal_bound_comp}. We compute reachable sets under each experimental condition from a restricted initial position set of $\mathcal{X}_0=[-0.51, -0.49]$. For each experiment, the verification for the initial set $\mathcal{X}_0$ was carried out in parallel with 200 sub-intervals of size $0.0001$. 
The average time to compute reach sets for each of the 200 initial subsets is shown in Table \ref{tab:verify_results}. 
For further evaluation, Figure~\ref{fig:reachtubes} provides a visual comparison between the reachable sets produced by our best method (GA + ETDL, $M=7$) and by the baseline for the entire initial set $\mathcal{X}_0=[-0.55, -0.45]$.
\begin{table}[t]
\tiny
\centering
\begin{adjustbox}{width=\textwidth}
\begin{tabular}{|l|c|c|c|c|c|c|c|c|c|c|c|c|c|c|}
    \hline
        &
        \multicolumn{7}{c|}{\textbf{Average Time to Compute 90 Steps [s]}} &
        \multicolumn{7}{c|}{\textbf{Max Reachable Set Size over 90 Steps}}\\
            \hline
        \diagbox[dir=NW]{\textbf{Algorithm}}{\vspace{-1mm}\\$M$\vspace{-1mm}}  
        & 1 & 2 & 3  & 4 & 5&6&7& 
        1 & 2 & 3  & 4 & 5&6&7 \\
        \hline

Uniform& 1,066 & 2,308 & 4,167 & 3,386 & 19,824 & 9,059 & 9,745 & 0.939 & 0.883 & 0.406 & 0.230 & 0.520 & 0.251 & 0.203\\
 \hline 
SA + EL & - & 2,356  & 2,674 & 3,855 & 5,318 & 6,021 & 7,195  & - & 0.458 & 0.225 & 0.226 & 0.213&  0.218& 0.210 \\
\hline
SA + ETDL & - & 2,401 & 2,075 & 2,371 & 2,963  & 4,054 & 4,737 & - & 0.448 & \textbf{0.200} & 0.205 & 0.165 & 0.164 & 0.162 \\

\hline
GA + EL & -& 2,490 & 2,871 & 3,970 & 9,580 & 12,231 & 9,601 & -&0.458 & 0.225 & 0.207 & 0.167 & 0.167 & 0.145\\
\hline
GA + ETDL &-& 2,405 & 2,084 & 2,261 & 2,729 & 3,323 & 5,249 &-& 0.456 & 0.203 & \textbf{0.168} & \textbf{0.163} & \textbf{0.154} & \textbf{0.115}\\
\hline
        \makecell{Time-based baseline \\ \cite{cleaveland2024conformal}} & \multicolumn{7}{c|}{\textbf{1,044}} &\multicolumn{7}{c|}{0.225}\\
        \hline
    \end{tabular}
    \end{adjustbox}
    \vspace{-2mm}
    \caption{Max reachable set size and 
    average 
    computation time 
    per initial subset
    for our state-based conformal bounds compared to the baseline time-series bounds from \cite{cleaveland2024conformal}.}
    \label{tab:verify_results}
    
    \vspace{-5mm}
\end{table}


\paragraph{Results \& Discussion.}
Overall, the genetic algorithm finds the smallest reachable set sizes. The most notable improvements come from our timed-decayed loss function ETDL, which greatly improves verification time and reduces reachable sets as compared to EL alone. This confirms the intuition that incurring error early in the verification process disproportionately impacts the resulting reachable sets. As compared to the baseline, our best-performing algorithm and loss combination GA+ETDL produces smaller reachable set sizes for all $M\geq 3$, though computing the reachable sets is much slower due to reachable sets potentially intersecting multiple regions at the same time, as noted in Section \ref{subsec:reachability}. 
See Appendix \ref{sec:OPT_appendix} for additional analysis of the subset merging optimization to handle this scalability challenge.

\section{Future Work and Conclusion}\label{sec:future}
In this paper, we presented a novel approach to finding state-dependent conformal bounds for a neural perception model and utilize knowledge of the system dynamics to produce high-confidence reachable sets via a symbolic verification tool. Our case study demonstrates our methods can produce dramatically tighter reachable sets than the state-of-the-art conformal method based on time series.

\looseness=-1
In future work, we plan to extend our methods to partition more state dimensions. These additional dimensions will motivate scalability improvements for optimization methods, data usage, and verification complexity. We also intend to investigate the case where the perception data is collected \textbf{off-policy}, i.e., using an exploratory controller. This would enable us to use an adaptive controller at run-time which may navigate dynamical environments with high confidence. Furthermore, we will investigate the \textbf{off-model} problem, where guarantees from simulations or a lab-tested system are extended to the deployed system. The combination of these methods would allow us to bridge the Sim2Real gap and enable the rapid and high-confidence development of safe autonomous systems.

\acks{This work was supported by the NSF Grants CCF-2403615 and CCF-2403616. Any opinions, findings,  conclusions, or recommendations expressed in this material are those of the authors and do not necessarily reflect the views of the National Science Foundation (NSF) or the US Government. }

\bibliography{bibFile.bib}

\newpage

\appendix
\section{Proofs}
\label{sec:proofs_appendix}

\subsection{Proof of Proposition~\ref{prop:sd_conformal_prediction}}
Consider the event $A = \{\exists k = 0..T: \|y_k - g(x_k)\| > \eta(x_k)\}$. Since the $\mathcal{S}_i$ are disjoint, we can bound the probability of $A$ as follows:
\begin{align*}
    \mathbb{P}_{x_0\sim \mathcal{D}_0, v_k \sim \mathcal{V}_{k|k-1}}[A] &= \mathbb{P}_{x_0\sim \mathcal{D}_0, v_k \sim \mathcal{V}_{k|k-1}}\left[\bigcup_{i=1}^M\{\exists k = 0..T: (\|y_k - g(x_k)\| > \eta(x_k) \wedge x_k \in \mathcal{S}_i)\}\right] \le \alpha,
\end{align*}
where the inequality follows from the union bound. 

\subsection{Proof of Theorem~\ref{thm:reachability}}
Consider the event $A = \{\exists k = 0..T: x_k \notin \mathcal{X}_k\}$. Since the sets $\mathcal{X}_i$ are worst-case reachable sets, then it must be the case that $\|v_l\| > \eta(x_l)$ for some $l \le k$., i.e., $$A = \{\exists k = 0..T: (x_k \notin \mathcal{X}_k \wedge \exists l \le k: \|v_l\| > \eta(x_l))\}.$$ However, we know that the noise bounds hold with probability $1-\alpha$ over the entire trajectory, so $\mathbb{P}_{x_0\sim \mathcal{D}_0, v_k \sim \mathcal{V}_{k|k-1}}[A] \le \alpha$, and the result follows.

\subsection{Proof of Proposition~\ref{prop:eta}}
Since each set $\Delta_i$ contains the maximal errors per trajectory within the corresponding region $\mathcal{S}_i$, the (normallized) $1-(\alpha/M)$ quantile provides a high-confidence bound on the trajectory-wide error within $\mathcal{S}_i$.

\section{Case Study Details}
\label{sec:MC_appendix}

\begin{figure}[h]


\subfigure[Canonical Environment][b]{%
\label{fig:mc_canonical}
\includegraphics[trim = 4 4 4 4, clip, width=.45\linewidth]{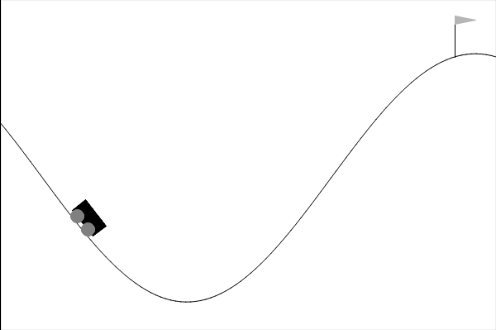}
}
\hfill
\subfigure[Low Contrast Environment][b]{%
\label{fig:mc_blurred}
\includegraphics[trim= 4 4 4 4 , clip, width=.45\linewidth]{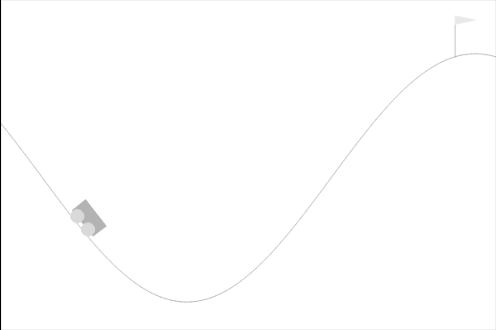}
}
\subfigure[High Contrast Environment][b]{%
\label{fig:mc_high_contrast}
\includegraphics[trim= 4 4 4 4 , clip, width=.45\linewidth]{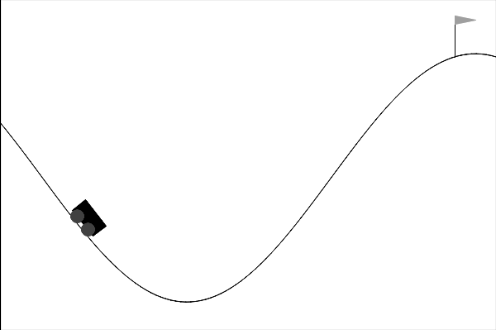}
}\hfill
\subfigure[Blurred Environment][b]{%
\label{fig:mc_low_contrast}
\includegraphics[trim= 4 4 4 4 , clip, width=.45\linewidth]{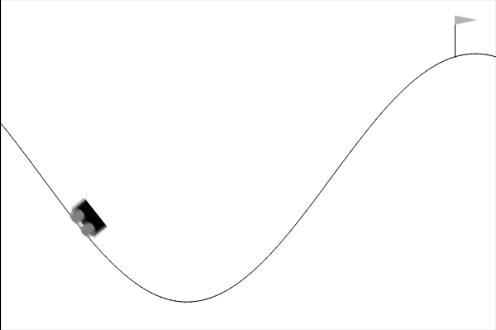}
}
\caption{Images of the mountain car environment}
\end{figure}

\paragraph{MC Background.}
Mountain Car is a common yet challenging reinforcement learning benchmark in which an underpowered car must reach the top of the right hill as shown in Figure \ref{fig:mc_canonical}. Because the car is underpowered, a successful controller must first utilize the left hill to gain momentum before reaching the goal on the right side. The car dynamics are shown in \eqref{eq:mc_dynamics} where $p\in[-1.2, 0.6]$ is the position, $v\in [-0.07, 0.07]$ is the velocity, and $u \in [-1.0, 1.0]$ is the control thrust. The initial position is the bottom of the mountain with $p_0 \in [-0.55, -0.45]$ and at rest with $v_0 = 0$. The dynamics for the standard system are as follows:
\begin{align}
\label{eq:mc_dynamics}
\begin{split}
    p_{k+1} &= p_k + v_k\\
    v_{k+1} &= v_k + 0.0015u_k -0.0025\cos{(3p_k)}\\
\end{split}
\end{align}

\paragraph{Case Study System \& Environment.}
For our case study, we use neural networks for both the state-based controller $h: p\times v\mapsto u$ and an image-based perception model $nn:z\mapsto\hat{p}$ that observes images $z$ of the MC environment and produces a state estimate $\hat{p}$. We use the simulator as the canonical perception map $s: p\mapsto z$ to map positions to $400\times600$ pixel gray-scaled images of the environment. Thus, our case study system is as follows: 
\begin{align}
\label{eq:mc_case_study_dynamics}
\begin{split}
    z_k &= s(p_k)\\
    \hat{p}_k &= nn(z_k) \\
    u_k &= h(\hat{p}_k, v_k) \\
    v_{k+1} &= v_k + 0.0015u_k -0.0025\cos{(3p_k)}\\
    p_{k+1} &= p_k + v_k\\
\end{split}
\end{align}
We add contrast and blur noise to images for training and deployment, respectively, of the perception model. In particular, we consider modified perception maps and noise parameters $\alpha$ and $\delta$. For contrast, the modified perception map $s_c:p\times\alpha \mapsto z_c$ creates a contrasted image $z_c$.  Contrast is added using the Python Image Library (PIL) ImageEnhance module where $\alpha=0$ produces a solid gray image, $\alpha=1$ produces the original image, and $\alpha>1$ produces a higher contrast version of the original image (\cite{clark2015pillow}).  The blur perception map $s_b:p\times\delta \mapsto z_b$ creates a blurred image $z_b$. Blur is added as follows: $z_{b} = s_b(p, \delta) = 0.5s(p-\delta) + s(p) +0.5s(p+\delta)$, i.e.,~a canonical image with a lighter overlay of left and right shifted images. Blurred images are then normalized to [0, 1].

\paragraph{Controller.}
The controller, $h$, is a neural network that takes position and velocity as inputs, has two hidden layers of 16 neurons with sigmoid activations, and has one output neuron with tanh activation. This controller was pre-trained and pre-verified to be safe by \cite{ivanov19}, i.e., it reaches the goal with a reward of at least 90 when starting in the initial set $p_0\in[-0.59, -0.4]$ when observing ground truth position and velocity. For this case study, we consider the initial set $p_0\in [-0.55, -0.45]$ for which the controller is more robust.

\paragraph{Perception Model.} The state estimator $nn$ is a convolutional neural network (CNN). The input is a single channel (gray-scaled) $400\times600$ pixel image followed by 2 (convolutional + max pooling) layers with 16 internal channels followed by 2 hidden linear layers or 100 neurons and a single output neuron. The convolutional layers have kernels of 32 and 24, and the pooling kernel is size 16. Stride is 2 for both convolution and pooling. All internal activations functions are ReLU, and the output is a scaled and shifted Tanh such that outputs are in the range of the MC position: $[-1.2, 0.6]$. The model is trained on contrasted and canonical images (see Figures \ref{fig:mc_canonical}-\ref{fig:mc_high_contrast}) generated from 100 equally-spaced positions and 9 contrast levels from $\alpha \in [0.1, 2.0]$ for a total of 1000 samples. The model was trained for 1000 epochs with MSE Loss. 

\paragraph{Trajectory Data Collection.} During data collection, we added out-of-distribution blur noise to images with $\delta=0.005$. See Figure \ref{fig:mc_blurred} for for an example image. 

\section{Reachability Computation Optimizations}
\label{sec:OPT_appendix}

\begin{table}[h]
\tiny
\centering
\begin{tabular}{|l|c|c|c|c|c|c|c|c|c|c|c|c|c|c|}
    \hline
        &
        \multicolumn{7}{|c|}{\textbf{Average Time to Compute 90 Steps [s]}} &
        \multicolumn{7}{|c|}{\textbf{Max Reachable Set Size over 90 Steps}}\\
            \hline
        \diagbox[dir=NW]{\textbf{Algo}}{\textbf{M}}  
        & 1 & 2 & 3  & 4 & 5&6&7& 
        1 & 2 & 3  & 4 & 5&6&7 \\
        \hline
 G + ETDL (Greedy Merge) & - & \textbf{1,471} & \textbf{1,815} & \textbf{1,617} & \textbf{1,819} & \textbf{2,130} & \textbf{3,096} & -& 0.606 & 0.287 & 0.246 & 0.237 & 0.219 & 0.154 \\
 \hline  
 G + ETDL (OPP Merge) & - & 2,405  & 2,084 & 2,261 & 2,729 & 3,323 & 5,249 & - & \textbf{0.456} & \textbf{0.203} & \textbf{0.168} & \textbf{0.163} & \textbf{0.154} & \textbf{0.115}\\
 \hline
 G + ETDL (No Merge) & - & 3,260 & 2,508 & 2,822 & 4,463 & 5,687 & 11,605 &  - & \textbf{0.456} & \textbf{0.203} & \textbf{0.168} & \textbf{0.163} & \textbf{0.154} & \textbf{0.115}\\
 \hline
    \end{tabular}
    \caption{Comparison of time to compute reachable sets and reachable sets sizes based for different subset merging algorithms.}
    \label{tab:verify_opt_results}
    
\end{table}
As described in Section \ref{subsec:reachability}, shrink-wrapping in the verification tool allows to remove or \emph{merge} redundant branches with existing branches opportunistically. Table \ref{tab:verify_opt_results} shows how this opportunistic method (OPP Merge) reduces verification time while not introducing additional over-approximation error as compared to not removing redundant branches  (No Merge). As an extension, we could additionally allow increased approximation error in the interest of time. One way is to greedily shrink-wrap branches whenever they \emph{would} have children to be removed. This method (Greedy Merge) is shown in Table \ref{tab:verify_opt_results}, and the results demonstrate the the corresponding time decrease and over-approximation increase. Future work will consider other optimizations for this scalability challenge, particularly as the number of regions increase with additional state dimensions.

\end{document}